\documentclass[fleqn,12pt,twoside]{article}
\usepackage[headings]{espcrc1}

\readRCS
$Id: espcrc1.tex,v 1.2 2004/02/24 11:22:11 spepping Exp $
\usepackage{graphicx}
\usepackage[figuresright]{rotating}

\newcommand{\AmS}{{\protect\the\textfont2
  A\kern-.1667em\lower.5ex\hbox{M}\kern-.125emS}}

\title{Dynamical Phase Trajectories in Baryon and Isospin Density Spaces}

\author{M. Colonna\address[Lab]{INFN-LNS,  
        via Santa Sofia 62, I-95123 Catania, Italy}, 
        V. Baran\address{NIPNE-HH, Bucharest and Bucharest University, 
        Romania},
        M. Di Toro\addressmark[Lab]\address
       {Physics and Astronomy Dept., University of Catania, Italy}
        and
        V. Giordano\addressmark[Lab]}
       
\begin{document}

\maketitle

\begin{abstract}
We review recent results obtained for charge asymmetric systems at Fermi
and intermediate energies, ranging from 30 MeV/u to 1 GeV/u.
Observables sensitive to the isospin dependent part of nuclear interaction
are discussed, providing information on the symmetry energy behavior from
sub- to  supra-saturation densities.   
\end{abstract}

\section{Introduction}

Nuclear reactions give us the opportunity to create transient states of nuclear
matter, 
following several dynamical paths in temperature, baryon and isospin density 
spaces. 
By looking at appropriate mechanisms, and related observables, along these trajectories, 
one can try to map the behavior of the nuclear interaction away from normal conditions.
 In particular, we want to investigate the energy
functional of asymmetric nuclear matter and constrain the term depending
on the asymmetry parameter $I = (N-Z)/A$, the so-called symmetry 
energy, $E_{sym}$, that is still largely debated nowadays \cite{baranPR,baoPR08}. 
Suitable parametrizations of the symmetry potential can be inserted into existing
transport codes (here we will follow the SMF approach \cite{chomazPR}), 
providing predictions for isospin-sensitive
observables, that can be confronted to experimental data. 
We stress that  
the knowledge of the Equation of State 
of asymmetric matter (Iso-EOS) 
has important implications in the context of structure studies
and astrophysical problems. 

At the Fermi energies, where one essentially explores the low-density zone of
the nuclear matter phase diagram, isospin effects can be investigated 
in 
reaction mechanisms typical of this energy domain, such as deep-inelastic
collisions and multifragmentation. 
The high density symmetry term can be probed from
 isospin effects appearing in heavy ion reactions
at relativistic energies (few GeV/u range).
Rather isospin sensitive observables are proposed from nucleon/cluster 
emissions, collective flows
and meson production. 
A large symmetry repulsion at high baryon density
will also lead to an ``earlier'' hadron-deconfinement transition
in n-rich matter. 

In the following, we will test  an $Asysoft$ parametrization of $E_{sym}$,
with an almost flat behavior below
 $\rho_0$ and even decreasing at supra-saturation, or an $Asystiff$ behavior, 
with a faster decrease at lower densities and much stiffer above saturation. 



\begin{figure}[h]
\includegraphics[width=20pc]{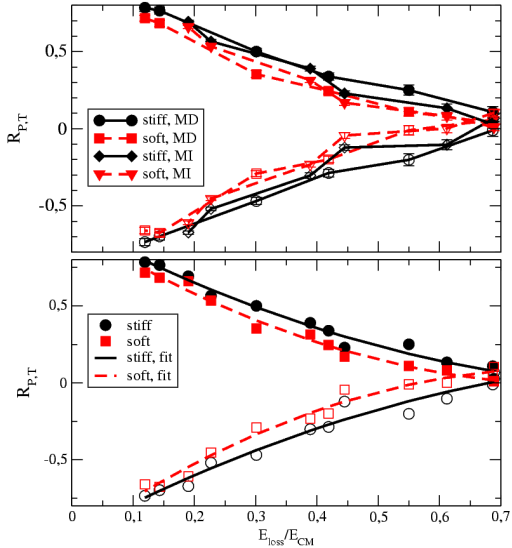}\hspace{2pc}%
\begin{minipage}[b]{14pc}\caption{\label{imb_eloss}
Imbalance ratios as a function of relative energy loss. 
Upper panel: separately for 
stiff (solid) and soft (dashed) Iso-EOS, and for 
two parametrizations of the isoscalar part of the interaction: 
MD 
(circles and squares) and MI 
(diamonds and triangles), in the projectile region (full symbols)
 and the target region 
(open symbols).
Lower panel: quadratic fit to all points for the stiff (solid), resp.
 soft (dashed) 
Iso-EOS.}
\end{minipage}
\end{figure}  

\section{Low density behavior of $E_{sym}$ : Isospin equilibration}

In this section we focus on the mechanisms connected to isospin transport in 
binary events at Fermi energies. 
This process involves nucleon exchange 
through the low density neck region and hence it is sensitive to
the low density behavior of $E_{sym}$ \cite{tsang92,isotr07,sherry}.  

Within a first order approximation of the transport dynamics, the relaxation
of a given observable $x$ towards its equilibrium value can be expressed as:
$x_{P,T}(t) - x^{eq} = (x^{P,T} -  x^{eq})~e^{-t/\tau}$,
where $x^{P,T}$ is the initial $x$ value for the projectile (P) or the target (T), 
$x_{eq} = (x^P + x^T)/2$ is the full equilibrium value, $t$ is the elapsed time
and $\tau$ is the relaxation time, that depends on the mechanism under study.   
The degree of isospin equilibration reached in the collision can be inferred by looking 
at isospin dependent observables in the exit channel, such as the N/Z of
PLF and TLF. 
Using the dissipated kinetic energy as a measure of the contact time $t$, 
one can finally extract the information on the relaxation time $\tau$,
that is related to the symmetry energy. 
It is rather convenient to construct the so-called imbalance ratio,
$R^x_{P,T} = {(x_{P,T}-x^{eq})} / {|x^{P,T}-x^{eq}|}~$ \cite{tsang92}.
Within our approximation, it simply reads: $R_{P,T} = \pm e^{-t/\tau}$.   
The simple arguments developed above are confirmed by full simulations of
(Sn,Sn) collisions at 35 and 50 MeV/u
\cite{isotr07}. 
In figure \ref{imb_eloss} we report the correlation between
$R_{P,T}$ 
and the total kinetic energy loss, that is used as a selector
of the reaction centrality and, hence, of the contact time $t$.
On the bottom part of the figure, where all results are collected together, one can see
that all the points essentially follow a given line,
depending only on the symmetry energy parametrization adopted. A larger
equilibration (smaller $R$) is observed in the $Asysoft$ case, corresponding to
the larger value of $E_{sym}$.
An experimental study of isospin diffusion as a function of the dissipated
kinetic energy  has been performed recently, by looking at the isotopic content
of the light charged particle emission as an indicator of the N/Z of the PLF \cite{Indra}.
This analysis points to a symmetry energy behavior in between the two
adopted parametrizations, in agreement with other recent estimates \cite{bettynew}.   

\begin{figure}[t]
\vskip 0.3cm
\includegraphics[width=20pc]{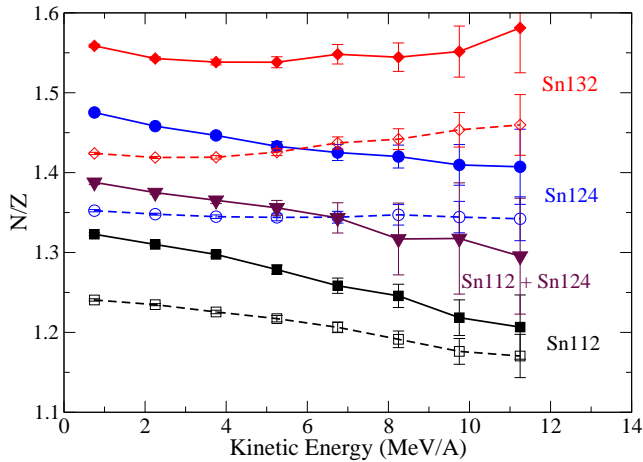}
\hskip 1.cm
\begin{minipage}[b]{14pc}\caption{ 
The fragment N/Z as a function of the kinetic energy for several (Sn,Sn) collisions
at E/A = 50 MeV/u and $b = 2~fm$. 
Full lines: $Asystiff$;  Dashed lines: $Asysoft$.
}
\end{minipage}
\label{iso_kin}
\end{figure}

\section{Isospin distillation in central collisions}

In central collisions at 30-50 MeV/u, where the full disassembly of the system
into many fragments is observed, one can study specifically properties of
liquid-gas phase transitions occurring in asymmetric matter 
\cite{baranPR,baoPR08,chomazPR}. 
For instance,
in neutron-rich matter, phase co-existence leads to
a different asymmetry
in the liquid and gaseous phase:  fragments (liquid) appear more symmetric
with respect to the initial matter, while light particles (gas) are
more neutron-rich. This sharing of the neutron excess optimizes the energy
balance and is ruled by the derivative of the symmetry energy with respect
to density. 
Recently we have proposed to investigate the correlations between the 
distillation mechanism and the underlying expansion dynamics of the fragmenting
system. In fact, in neutron(poor)-rich systems, neutrons(protons) are more 
repulsed than protons(neutrons), building interesting correlations between the
fragment $N/Z$ and kinetic energy, that are sensitive to the symmetry energy
parametrization adopted in the calculations, see figure 2. 
As one can see in the figure, larger (negative) slopes are obtained in the $Asystiff$ case,
corresponding to the lower value of the symmetry energy at low density.  
This appears as a promising experimental observable to be investigated, though
fragment secondary effects are expected to reduce the sensitivity to the
Iso-EOS \cite {col07}.


\vskip -1.0cm
\section{Isospin effects at high baryon density}

\begin{figure}
\includegraphics[width=8.5cm]{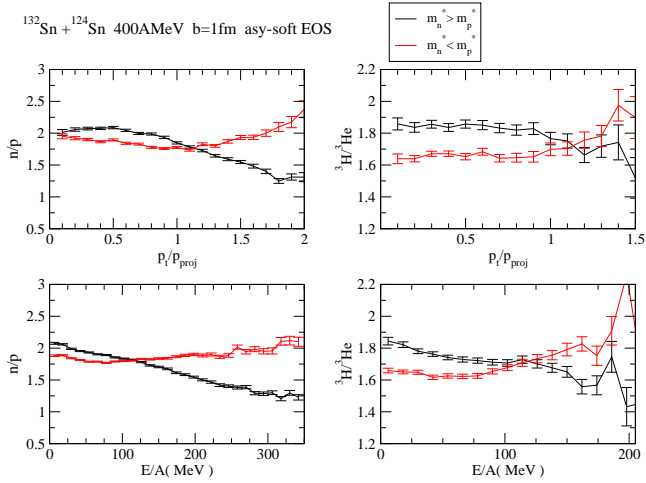}
\hskip 0.3cm
\begin{minipage}[b]{14pc}
\caption{$^{132}$Sn+$^{124}$Sn at 400 MeV/u, central collisions.  
Isospin content of 
nucleon and light ion emissions vs $p_t$ (upper) and kinetic energy (lower).
Results are shown for the $Asysoft$ interaction}.
\end{minipage}
\label{fastratios}
\end{figure}

The problem of Momentum Dependence in the Isovector
channel ($Iso-MD$) of the nuclear interaction (leading to  
neutron/proton effective mass splitting)
is still very controversial and it would be extremely
important to get more definite experimental information
\cite{BaoNPA735,rizzoPRC72}. 
Exotic Beams at intermediate energies are
of interest in order to have high momentum particles and to test regions
of high baryon (isoscalar) and isospin (isovector) density during the
reaction dynamics.
We present here some results for reactions induced by $^{132}Sn$
 beams on $^{124}Sn$ targets at 400 MeV/u \cite{vale08}. For central 
collisions in the interacting zone we can reach baryon densities about
$1.7-1.8~ \rho_0$ in a transient time of the order of 15-20 fm/c. 
In figure 3 we show the  $(n/p)$ and $^3H/^3He$ yield ratios at freeze-out,
for two choices of mass splitting, vs. transverse
momentum $p_T$ (upper curves) and kinetic energy (lower curves). In this way we can
separate particle emission from sources at different densities.
We note a clear decreasing trend only in the case $m^*_p < m^*_n$, 
corresponding to a larger proton repulsion. 
Similar results are obtained for $Asysoft$ or $Asystiff$ parametrizations.
Hence these data seem to be suitable to disentangle $Iso-MD$ effects,
rather than the stiffness of the symmetry energy.
An interesting dependence on the effective mass splitting is observed also
for other observables, such as collective flows, that are also sensitive to 
the $Asy$-stiffness at high density \cite{vale08}.

\section{Perspectives}

We have reviewed some aspects of the phenomenology associated with nuclear 
reactions, from which new hints
are emerging to constrain the EOS of asymmetric matter.  
The greatest theoretical uncertainties concerns the high density 
domain, 
that has the largest impact on the
understanding of the properties of neutron stars. 
In the near future, thanks to the availability of both stable and rare
isotope beams, more selective analyses, also based on new exclusive 
observables, 
are expected to provide further stringent constraints.

\end{document}